\documentclass[journal]{IEEEtran}
\usepackage{multirow}

\usepackage[noadjust]{cite}
\usepackage{amsmath}
\usepackage{amssymb}
\usepackage{amsthm}
\usepackage{booktabs} 
\usepackage{xcolor}
\usepackage{amssymb}
\usepackage{graphicx}
\usepackage{caption}
\usepackage{changepage}
\usepackage{adjustbox}
\usepackage{scrextend}
\usepackage{makecell}
\usepackage{url}
\usepackage{comment}
\usepackage[vlined, ruled]{algorithm2e} 
\usepackage{enumitem}
\usepackage{tablefootnote}

\newcommand{\deepm}{\textsc{DeepMalware}\xspace}
\newcommand{\spectrum}{\textsc{Propedeutica}\xspace}

\newcommand{\hybridm}{\textsc{Propedeutica}\xspace}
\newcommand{\reconm}{\textsc{Reconstruction Module}\xspace}

\newcommand{\xy}[1]{{\color{black} #1}\normalfont}
\newcommand{\xyn}[1]{{\color{black} #1}\normalfont}
\newcommand{\yj}[1]{{\color{black} #1}\normalfont}

\begin{document}

\title{Learning Fast and Slow: \spectrum for Real-time Malware Detection}
\author{\IEEEauthorblockN{Ruimin Sun\IEEEauthorrefmark{2}\IEEEauthorrefmark{1}, Xiaoyong Yuan\IEEEauthorrefmark{3}\IEEEauthorrefmark{1}, Pan He\IEEEauthorrefmark{4}, Qile Zhu\IEEEauthorrefmark{4}, Aokun Chen\IEEEauthorrefmark{4}, Andre Gregio\IEEEauthorrefmark{5},\\ Daniela Oliveira\IEEEauthorrefmark{4}, and Xiaolin Li\IEEEauthorrefmark{9}}\\
\IEEEauthorblockA{\IEEEauthorrefmark{2}Northeastern University}
\IEEEauthorblockA{\IEEEauthorrefmark{3}Michigan Technological University}
\IEEEauthorblockA{\IEEEauthorrefmark{4}University of Florida}\\
\IEEEauthorblockA{\IEEEauthorrefmark{5}Federal University of Parana}
\IEEEauthorblockA{\IEEEauthorrefmark{9}Cognization Lab}\\
\IEEEauthorblockA{\IEEEauthorrefmark{1}Equal Contribution}
}


\maketitle

\begin{abstract}
Existing malware detectors on safety-critical devices have difficulties in runtime detection due to the performance overhead. In this paper, we introduce \spectrum\footnotemark, a framework for efficient and effective real-time malware detection, leveraging the best of conventional machine learning (ML) and deep learning (DL) techniques. In \spectrum, all software start execution are considered as benign and monitored by a conventional ML classifier for fast detection. If the software receives a borderline classification from the ML detector (e.g. the software is 50\% likely to be benign and 50\% likely to be malicious), the software will be transferred to a more accurate, yet performance demanding DL detector. To address spatial-temporal dynamics and software execution heterogeneity, we introduce a novel DL architecture (\deepm) for \spectrum with multi-stream inputs. We evaluated \spectrum with 9,115 malware samples and 1,338 benign software from various categories for the Windows OS. 
With a borderline interval of [30\%-70\%], \spectrum achieves an accuracy of 94.34\% and a false-positive rate of 8.75\%, with 41.45\% of the samples moved for \deepm analysis. Even using only CPU, \spectrum can detect malware within less than 0.1 seconds.

\end{abstract}

\IEEEpeerreviewmaketitle

\footnotetext{\textit{Propedeutics} refers to diagnosing a patient condition by first performing initial non-specialized, low-cost exams, and then proceeding to specialized, possibly expensive diagnostic procedures if preliminary exams are inconclusive.}

\section{Introduction} \label{sec:intro}
Malware is continuously evolving~\cite{callejo16}, and existing protection mechanisms have not been coping with their sophistication~\cite{fratantonio2016triggerscope}. 
The industry still heavily relies on signature-based technology for malware detection~\cite{vigna98}, but these methods have many limitations: (i) they are effective only for malware with known signatures; (ii) they are not sustainable, given the massive amount of samples released daily; and (iii) they can be evaded by zero-day and polymorphic/metamorphic malware (practical detection 25\%-50\%)~\cite{bromium}. 

Behavior-based approaches attempt to identify malware behaviors using instruction sequences~\cite{semantics-aware,miningspec}, computation trace logic~\cite{kinder05}, and system or API call sequences \cite{kolbitsch09,canali12,bayer2006ttanalyze}. These solutions have been mostly based on conventional machine learning (ML) models,
such as K-nearest neighbor, SVM, neural networks, and decision tree algorithms~\cite{hofmeyr1998intrusion,bayer2009scalable,revathi2013detailed,rieck2011automatic}. However, current solutions based on ML still suffer from high false-positive rates, mainly because of (i) the complexity and diversity of modern benign software and malware~\cite{lanzi2010accessminer,canali12,paloalto,callejo16}, which are hard to capture during the learning phase of the algorithms; (ii) sub-optimal feature extraction; (iii) limited training/testing datasets, and (iv) the challenge of concept drift~\cite{widmer1996learning}, which makes it hard to generalize learning models to reflect future malware behavior. 

The accuracy of malware classification depends on gaining sufficient context information about software execution and on extracting meaningful abstraction of software behavior. For system/API-call based malware classification, longer sequences of calls likely contain more information. However, conventional ML-based detectors (e.g., Random Forest~\cite{breiman2001random}) often use short windows of system calls during the training phase to avoid the curse of dimensionality (when the dimension increases, the classification needs more data to support and becomes harder to solve~\cite{bengio2007scaling}), and may not be able to extract useful features for accurate detection.
Thus, the main drawback of current behavioral-based approaches is that they might lead to low accuracy and many false-positives because it is hard to analyze complex and longer sequences of malicious behaviors with limited window sizes, especially when malicious and benign behaviors are interposed. For instance, the n-grams of system calls are widely used as features of ML detection, which, however, are prone to bring blind spots in the detection.

In contrast, emerging deep learning (DL)
models~\cite{lecun2015deep} are capable of analyzing longer sequences of system calls and making more accurate classification through higher level information extraction, while circumventing the curse of dimensionality. \xyn{However, DL requires more time to gain enough information for classification and to predict the probability of detection. Moreover, the state-of-the-art DL models consist of deep layers and a huge amount of parameters, which results in slow calculation in the prediction. Such slow process makes DL models infeasible to provide predictions on malware in real time. }
This trade-off is challenging: fast and perhaps not-so-accurate (ML methods) \textit{vs.} time-consuming and more accurate classification (DL methods). \xy{Applying a single method only can be  either inefficient or ineffective, which becomes impractical for real-time malware detection. }

\xy{In this paper, we introduce and evaluate \spectrum, a framework for efficient and effective real-time malware detection, considering both prediction performance and efficacy in the detection phase. \spectrum is designed to combine the best of ML and DL methods, i.e., speed in prediction from the ML methods and accuracy from the DL methods.}  In \spectrum, all software in the system is subjected to ML for fast classification. If a piece of software receives a borderline malware classification probability, it is then subjected to additional analysis with a more performance expensive and more accurate DL classification. The DL classification is performed via our proposed algorithm, \deepm. Compared to existing DL methods~\cite{hardy2016dl4md,wang2017adversary,kolosnjaji2016deep}, \deepm can learn spatially local and long-term features and handle software heterogeneity via the application of both ResNext blocks and recurrent neural networks with multi-stream inputs.
By leveraging such multi-stage design, \spectrum is capable of providing precise and real-time detection. 


We evaluated \spectrum with a set of 9,115 malware samples and 1,338 common benign software from various categories for the Windows OS. \spectrum leveraged sliding windows of system calls for malware detection. Our proposed deep learning algorithm, \deepm, achieved a 97.03\% accuracy and a 2.43\% false positive rate.
By combining conventional ML and DL, \spectrum substantially improves the prediction performance while keeping the detection time less than 0.1 seconds, which makes DL feasible for the real-time malware detection.

In this paper, we present the following contributions: (i) \spectrum, a new framework for efficient and effective real-time malware detection for Windows OS, (ii) an evaluation of \spectrum with a collection of 9,115 malware and 1,338 benign software, and (iii) \deepm, a novel DL algorithm, learning multi-scale spatial-temporal system call features with multi-stream inputs that can handle software heterogeneity.


\section{Methodology}
\subsection{Threat Model} \label{sec:threat}
\spectrum is designed to provide precise, real-time malware detection, i.e., a high accuracy rate, few false-positives, and less processing time.
\spectrum is capable of defending against the most prevalent
malware threats, and not against directed attacks, such as specifically-crafted, advanced, and persistent threats. Hence, we assume that malware will eventually get in if an organization is targeted by a well-motivated attacker. \xyn{Therefore, attacks either targeting \spectrum's
software implementation, or the machine
learning engine (adversarial machine
learning~\cite{yuan2019adversarial,zhang2019adversarial}) are outside of \spectrum's scope. We discuss the adversarial attacks against malware detection in Section~\ref{sec:adversarial}.}

\spectrum operation assumes that, before it loads, the system
is on a pristine state. Thus, \spectrum will not scan all running processes at startup, which limits its checking procedure to each new process created.


\vspace{-0.5em}
\subsection{Overall Design} \label{sec:arch}

\begin{figure}[!tb]
\centering
\includegraphics[width=\linewidth]{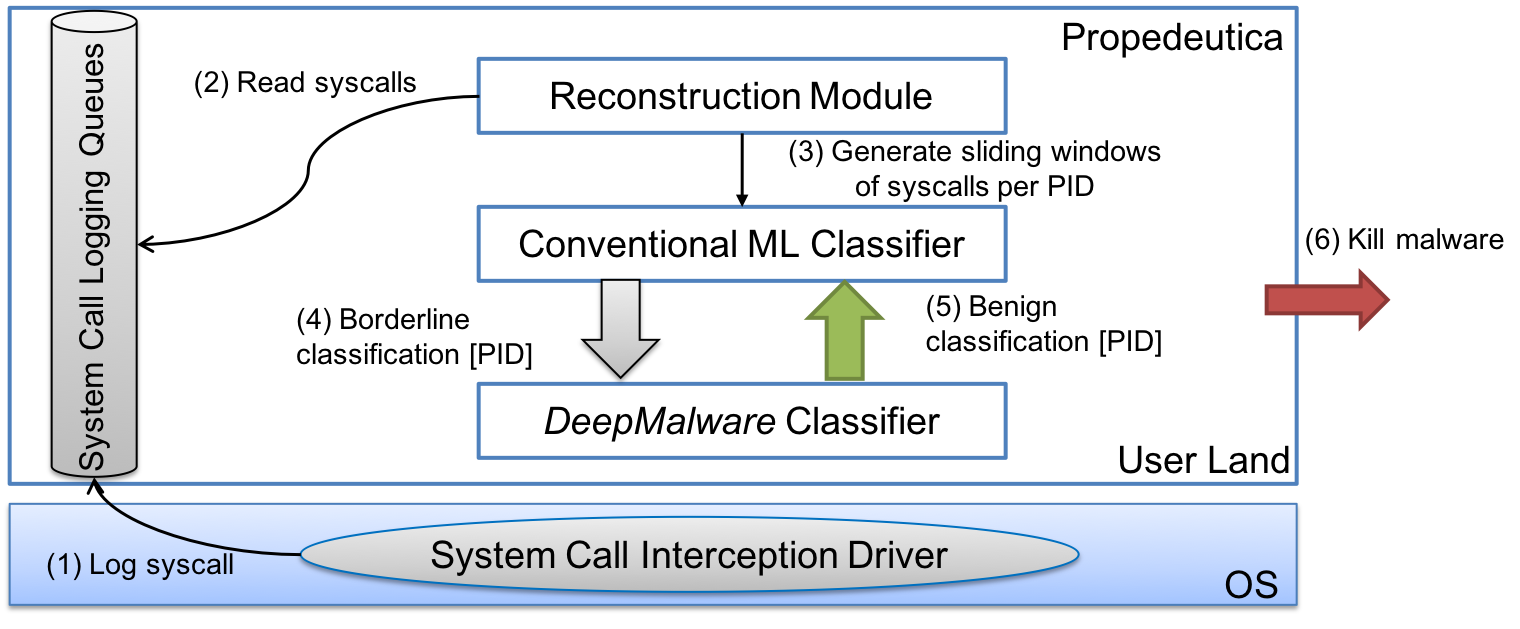}
\vspace{-1.5em}
\caption{Architecture and workflow of \spectrum for multi-stage malware detection.}
\label{fig:main1}
\vspace{-1.5em}
\end{figure}

\spectrum (Figure~\ref{fig:main1}) consists of: (i) a system call reconstruction module, (ii) an ML classifier, and (iii) our newly proposed \deepm classifier. 

The workflow of \spectrum is outlined as follows. 
\par\noindent
\textbf{STEP 1:} \spectrum leverages a System Call Interception Driver to intercept and log all system calls invoked in the system, and associate them with the PID of the invoking process.
\vspace{0.2em}\par\noindent
\textbf{STEP 2:} The Reconstruction Module reads and parses logged system calls, compressing the system calls for input to the two classifiers.
\vspace{0.2em}\par\noindent
\textbf{STEP 3:} The classifiers generate sliding windows of system call traces with the PID of the invoking process. 
\vspace{0.2em}\par\noindent
\textbf{STEP 4:} Conventional ML Classifier introduces a configurable borderline probability interval [lower bound, upper bound], which determines the range of classification that is considered inconclusive for detection. For example, let's consider that the borderline interval is [30\%-70\%]. If the ML classifier assigns the label ``malware'' for a piece of software whose classification range is within the predefined borderline interval, \spectrum considers its classification result inconclusive. If the software receives a ``malware'' classification probability less than 30\%, \spectrum considers that the software is not malicious. If the classification probability is greater than 70\%, \spectrum considers the software malicious. For the inconclusive case ([30\%-70\%] interval), the software continues to run, but is subjected to analysis by DeepMalware Classifier for a definitive classification. 
\vspace{0.2em}\par\noindent
\textbf{STEP 5:} If the software is classified as benign by DeepMalware Classifier, it continues running under the monitoring of the Conventional ML classifier.
\vspace{0.2em}\par\noindent
\textbf{STEP 6:} If the software is considered malicious, \spectrum kills the malicious software.
\vspace{5pt}
\xyn{By leveraging the fast speed in ML methods and high accuracy in DL methods, \spectrum aims to achieve precise and real-time detection. The main goal of this paper using \spectrum is to present the concept of a multi-stage malware detector that combines two distinct classification models---ML and DL. } 
Therefore, any 
presented implementation should be
understood as a proof-of-concept (PoC) to accomplish
such a goal, and not as a 
platform-specific solution. We intend that this PoC leads to a \spectrum version 
that will be implemented in the future by OS 
vendors to be integrated into their products.
Moreover, \spectrum does not rely on a GPU-powered system to speed up its detection procedures, which broaden the application of \spectrum.

For the sake of simplicity, our PoC was implemented in user land, i.e., as a user process in a non-privileged ring of execution. Thus, \spectrum's trusted computing base includes the OS kernel, the learning models running in user land, and the hardware stack. 
We modeled \spectrum's PoC for Microsoft Windows, since it is the most targeted
OS by malware writers~\cite{w7}. We limited our PoC to the 32-bit Microsoft Windows version, which currently is the most popular Windows architecture.

\vspace{-0.5em}
\section{Implementation Details}
\label{sec:impl}
This section discusses implementation details of three main aspects of \spectrum's operation: the system call interception driver, the reconstruction module, and our newly proposed \deepm algorithm.

\vspace{-0.5em}
\subsection{The System Call Interception Driver}\label{sec:spectrum}

Intercepting system calls is a key step for \spectrum operation as it allows
collecting system call invocation data to be input to ML and DL classifiers. 
\yj{Despite many existing tools for process monitoring, such as Process Monitor~\cite{procmon}, drstrace library~\cite{drstrace}, Core OS Tool~\cite{coreostool}, and WinDbg's Logger~\cite{windbglogger}, they suffer from various challenges: Process Monitor provides only coarse-grained file and registry activity. drstrace works at the application level, and can be bypassed by malware. WinDbg's Logger starts logging only at the entry point of the execution, so system calls executed by the initialization code in statically imported shared libraries will not be seen. Core OS tools trace coarse-grained events, such as interrupts, memory events, thread creation and termination, and etc.
Therefore, we opted to implement our own solution to have more flexibility for deploying a real-time detection mechanism.} \spectrum collection driver was implemented for Windows 7 SP1 32-bit.  The driver hooks into the System Service
Dispatch Table (SSDT), which contains an array of function pointers to important 
system service routines, including system call routines. In Windows 7 32-bit system, there are 400 entries of system calls~\cite{ntapi}, of which 155 system calls 
are interposed by our driver. We selected a comprehensive set of system calls to
be able to monitor multiple, distinct subsystems, which includes network-related system calls (e.g., \textit{NtListenPort} and \textit{NtAlpcConnectPort}), file-related system calls (e.g., \textit{NtCreateFile} and \textit{NtReadFile}), memory-related system calls (e.g., \textit{NtReadVirtualMemory} and \textit{NtWriteVirtualMemory}), process-related system calls (e.g., \textit{NtTerminateProcess} and \textit{NtResumeProcess}), and other types (e.g., \textit{NtOpenSemaphore} and \textit{NtGetPlugPlayEvent}). 

\yj{This system call interception driver is publicly 
available at \cite{ourdriver}. 
We open sourced our system call interceptor so that future studies can generate real-time software execution traces, and evaluate the performance of their ML/DL algorithms in malware detection. 
Our solution can also be used as a baseline for comparison in future work. }

System call logs (timestamp, PID, syscall) 
were collected using DbgPrint 
Logger~\cite{dbgprintlog}, which enables 
real-time kernel message logging to a 
specified IP address (localhost or remote IP), 
thus allowing the logging pool and the 
\spectrum to reside on different hosts 
for scalability and performance.
The driver monitors all processes added
to the \textit{Borderline\_PIDs} list,
which keeps track of processes which 
received borderline classification from 
the conventional ML classifier. 
\xy{The time to collect system calls varies among different software. 
The time depends on the functionalities of the software (benign or malicious), and the types of system calls invoked, and the frequency of system call invocations. Hence, the time to collect system calls is software dependent.}

\vspace{-0.5em}
\subsection{The Reconstruction Module}\label{sec:recon}
In \spectrum, both the ML and the \deepm classifiers use the \reconm to preprocess the input system call sequences and obtain the same preprocessed data. 
The \reconm splits system calls sequences according to the PIDs of processes invoking them and parses them into three types of sequential data: process-wide n-gram sequence, process-wide density sequence, and system-wide frequency feature vector. Then, the \reconm converts the sequential data into windows using the sliding window method, which is usually used to translate a sequential classification problem into a classical one for every sliding window~\cite{forrest96}.

\vspace{0.3em}
\noindent\textbf{Process-wide n-gram sequence and density sequence.} We use the n-gram model, widely used in natural language processing (NLP) applications~\cite{manning1999foundations}, to compress system call sequences. N-gram is defined as a combination of $n$ contiguous system calls. The n-gram model encodes low-level features with simplicity and scalability and compresses the data by reducing the length of each sequence with encoded information. The process activity in a Windows system can be intensive depending on the workload, e.g., more than 1,000 system calls per second, resulting in very large sliding windows. Therefore, for \spectrum, we decided to further compress the system call sequences and translate them into two-stream sequences: n-gram sequences and density sequences. n-gram sequence is a list of n-gram units, while density sequence is the corresponding frequency statistics of repeated n-gram units. There are many possible n-gram variants such as n-tuple, n-bag, and other non-trivial and hierarchical combinations (e.g., tuples of bags, bags of bags, and bags of grams)~\cite{canali12}. \spectrum uses 2-gram because (i) compared with n-bag and n-tuple, the n-gram model is considered the most appropriate for malware system call-based classification~\cite{canali12} and (ii) the embedding layer and the first few convolutional neural layers can automatically extract high-level features from neighbor grams, which can be seen as hierarchical combinations of multiple n-grams. 
\xy{Once n-gram sequences fill up a sliding window, the \reconm delivers the window of sequences to ML classifier and then DL classifier (if in the inconclusive case) redirects the incoming system calls to the new n-gram sequences.}

\vspace{0.3em}
\noindent\textbf{System-wide frequency feature vector.}
Our learning models leverage system calls as features from all processes running in the system. This holistic, opposite to process-specific approach is more effective for malware detection compared to current approaches because modern malware can be multi-threaded~\cite{yin2007panorama,caillat2015prison}. System-wide information helps the models learn the interactions among all processes in the system. To gain whole system information, the \reconm collects the frequency of different types of n-grams from all processes during the sliding window and extracts them as a frequency feature vector. Each element of the vector represents the frequency of an n-gram unit in the system call sequence.

\vspace{-0.5em}
\subsection{\deepm Classifier}\label{sec:deep}

\spectrum aims to address spatial-temporal dynamics and software execution heterogeneity. Spatially, advanced malware leverages multiple processes to work together to achieve a long-term common goal. Temporally, a malicious process may demonstrate different types of behaviors (benign or malicious) during the lifetime of execution, such as keep dormant or benign at the beginning of the execution. 
To thoroughly consider available behavior data in space and time, we introduced a novel DL model, \deepm. 
\xy{First, to capture the spatial connections between multiple processes, we feed both process-wide features and system-wide features into \deepm. The process-wide and system-wide inputs provide detailed information about how the target software (malware or benign software) interacts with the rest software in the system. Second, to capture the temporal features, we first introduced the bi-directional LSTM~\cite{hochreiter1997long} to capture the connection of n-grams. However, due to the gradient vanishing problem of LSTM in processing long sequences, we further introduce multi-scale convolutional neural networks to extract high-level representation of n-gram system calls so as to reduce the input length of LSTM and avoid blind spots.}
To facilitate harmonious coordination, \deepm stacks spatial and temporal models for joint training.

\begin{figure*}[!ht]
\centering
\includegraphics[width=0.8\textwidth]{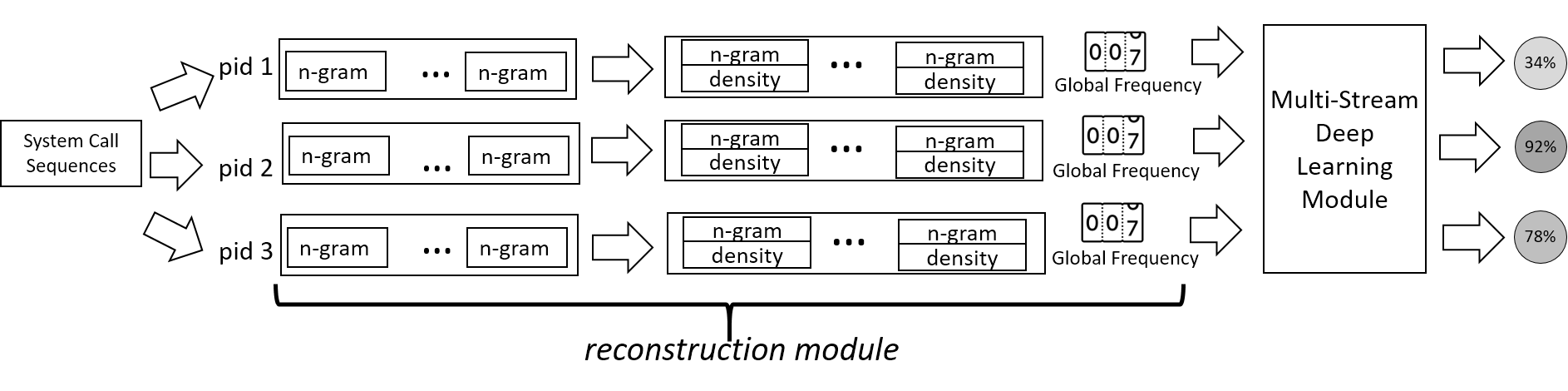}
\vspace{-0.5em}
\caption{Workflow of \deepm classification approach.}
\label{fig:dl_overview}
\vspace{-1.5em}
\end{figure*}

\begin{figure}[!ht]
\centering \includegraphics[width=0.95\linewidth]{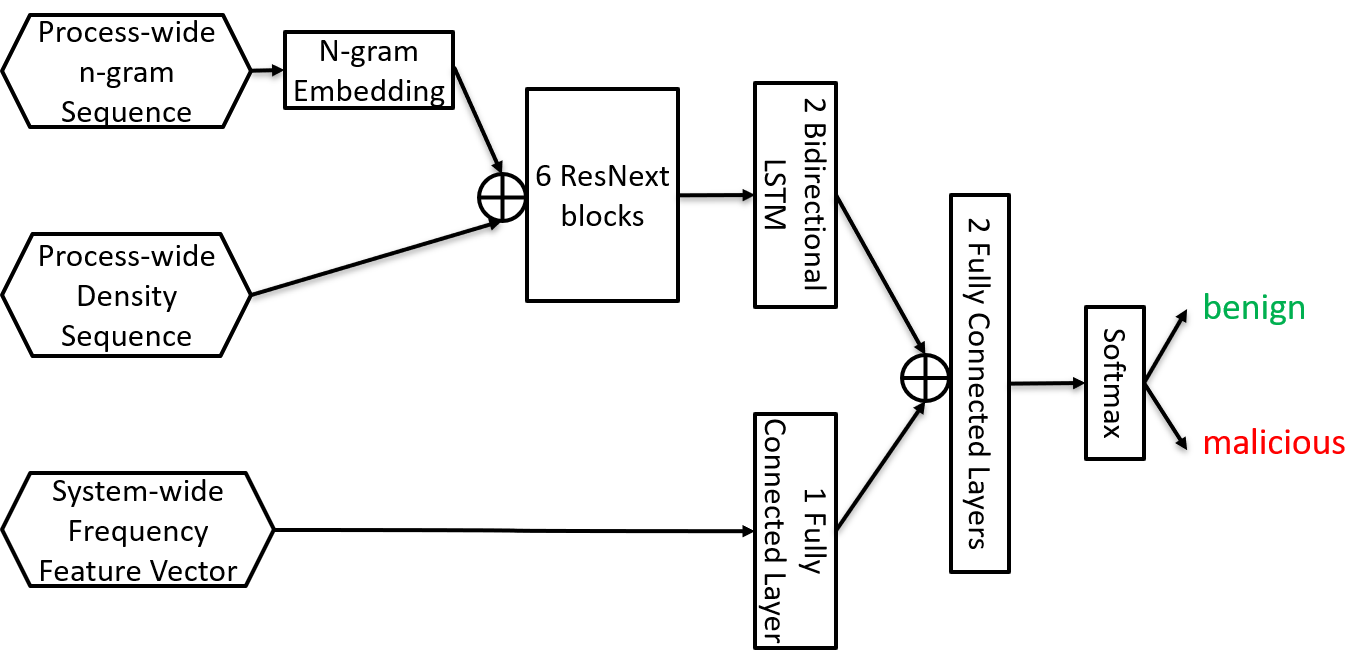}
\vspace{-0.5em}
\caption{\deepm architecture.}
\label{fig:dl_model}
\vspace{-1.5em}
\end{figure}

System call sequences are taken as input for all processes subjected to \deepm analysis (see Figure~\ref{fig:dl_overview} for a workflow of the classification approach). 
\deepm leverages n-gram sequences of processes and frequency feature vectors of the system. First, two streams (process-wide n-gram sequence and density sequence) model the sequence and density of n-gram system calls of the process. The third stream represents the global frequency feature vector of the whole system. \deepm consists of four main components: (i) N-gram Embedding, (ii) ResNext blocks, (iii) Long Short-Term Memory (LSTM) Layers, and (iv) Fully Connected Layers (Figure~\ref{fig:dl_model}).

\vspace{0.3em}
\noindent\textbf{N-gram Embedding.}
\deepm adopts an encoding scheme called N-gram Embedding, which converts sparse n-gram units into a dense representation. \deepm treats n-gram units as discrete atomic inputs. N-gram Embedding helps to understand the relationship between functionally correlated system calls and provides meaningful information of each n-gram to the learning system. Unlike vector representations such as one-hot vectors (which may cause data sparsity), N-gram Embedding mitigates sparsity and reduces the dimension of input n-gram units. The embedding layer maps sparse system call sequences to a dense vector. This layer also helps to extract semantic knowledge from low-level features (system call sequences) and largely reduces feature dimension. The embedding layer uses 256 neurons, which reduce the dimension of the n-gram model (number of unique n-grams in the evaluation) from 3,526 to 256.

\vspace{0.3em}
\noindent\textbf{ResNext blocks.} \deepm uses six ResNext blocks to extract different sizes of n-gram system calls in the sliding window. Conventional sliding window methods leverage small windows of system call sequences and, therefore, experience challenges when modeling long sequences. These conventional methods represent features in a rather simple way (e.g., only counting the total number of system calls in the windows without local information of each system calls), which may be inadequate for a classification task. 
The design of ResNext follows the implementation in~\cite{xie2017aggregated}. Each ResNext block aggregates multiple sets of convolutional neural networks with the same topology. ResNext blocks extract different n-grams to avoid blind spots in detection. Residual blocks perform short cuts between blocks to improve the training of deep layers. Batch normalization is applied to speed up the training process after convolutional layers and a non-linear activation function, ReLU to avoid saturation.

\vspace{0.3em}
\noindent\textbf{Long Short-Term Memory (LSTM) Layers.}
The internal dependencies between system calls include meaningful context information or unknown patterns for identifying malicious behaviors. To leverage this information, \deepm adopts one of the recurrent architectures, LSTM~\cite{hochreiter1997long}, for its strong capability of learning meaningful temporal/sequential patterns in a sequence and reflecting internal state changes modulated by the recurrent weights. LSTM networks can avoid gradient vanishing and learn long-term dependencies by using forget gates. LSTM layers gather information from the first two streams: process-wide n-gram sequence and density sequence.

\vspace{0.3em}
\noindent\textbf{Fully Connected Layers.}
\deepm deploys a fully connected layer is deployed to encode system-wide frequency. Then, it is concatenated with the output of bi-directional LSTMs to gather both sequence-based process information and frequency-based system-wide information. The last fully-connected layer with the softmax function outputs the prediction with probabilities.

\xy{Moreover, we adopt three popular techniques to avoid overtraining in \deepm. 1) We add a Dropout layer~\cite{srivastava2014dropout} with a dropout rate of 0.5 following the fully connected layer; 2) We include batch normalization~\cite{ioffe2015batch} in the ResNext blocks after the convolutional layers; 3) We use 20\% training data as a validation set and apply early stopping when the loss on the validation set does not decrease.}

\vspace{-0.5em}
\xy{\subsection{Conventional ML Classifier}\label{sec:ml}
We consider three pervasively used conventional ML algorithms: Random Forest (RF), eXtreme Gradient Boosting (XGBoost)~\cite{chen2016xgboost}, and Adaptive Boosting (AdaBoost)~\cite{freund1995desicion}. 
Random Forest and boosted trees have been considered as the best supervised models on high-dimensional data~\cite{caruana2008empirical}. 
We use AdaBoost and XGBoost as representatives of boosted trees. 

To select the best features used in machine learning classifiers, we consider both efficacy and effectiveness. In \spectrum, we take the frequency of n-gram units in a sequence as input features and estimate the probabilities of a sequence of system calls invoked by a malware. Specifically, a feature vector will be created for each sequence, and each element in the vector represents the frequency of the corresponding n-gram unit. We discuss the details of ML classifiers in Section~\ref{sec:exp}.

Notice that ML classifiers in \spectrum can be generalized to other ML algorithms, when ML classifiers can output classification probabilities for borderline classification (i.e., probabilistic ML models). 
For Non-probabilistic ML classifiers, we will adopt a conversion function to provide class probabilities. For example, the classification scores provided by Support Vector Machine can be mapped into the class probabilities via a logistic transformation with trained parameters~\cite{platt99probabilistic}. K Nearest Neighbor (kNN) methods can be extended with probabilistic methods for estimating the posterior probabilities using distances from neighbors~\cite{holmes2002probabilistic}. The classification probabilities of ML classifiers are used for the borderline classification (Fig.~\ref{fig:main1} Step 4). \xy{It is worth to note that \spectrum is designed to integrate with any machine learning and deep learning methods in malware detection with minimal efforts. Therefore, we select commonly used and representative algorithms to show \spectrum's capability in plugging in these algorithms.}}

\section{Evaluation}\label{sec:exp}
The main goal of our evaluation is to determine \spectrum's effectiveness for real-time malware detection. More specifically, we sought to discover to what extent \spectrum outperforms ML on prediction performance and DL on classification time.  


We first describe the dataset used in our evaluation. Next, we evaluate the performance of our newly proposed \deepm algorithm in comparison with the most relevant ML classifiers from the literature. Then, we present the results of our experiments on the proposed \spectrum.
In our evaluation, we compare the performance of classifiers when operating on a CPU and a GPU. 
GPUs not only reduce the training time for DL models but also help these models achieve classification time several orders of magnitude less than conventional CPUs in the deployment. 

We performed our test on a server running Ubuntu 12.04 with 2 CPUs, 4GB Memory, 60GB Disk. 
\deepm was implemented using Pytorch~\cite{pytorch}. The training and testing procedures ran on Nvidia Tesla M40 GPUs.

\subsection{Software Collection Method}
\begin{figure}[!t]
\centering
\includegraphics[width=0.7\linewidth,trim={0.1cm 0.1cm 0.1cm 0.1cm},clip]{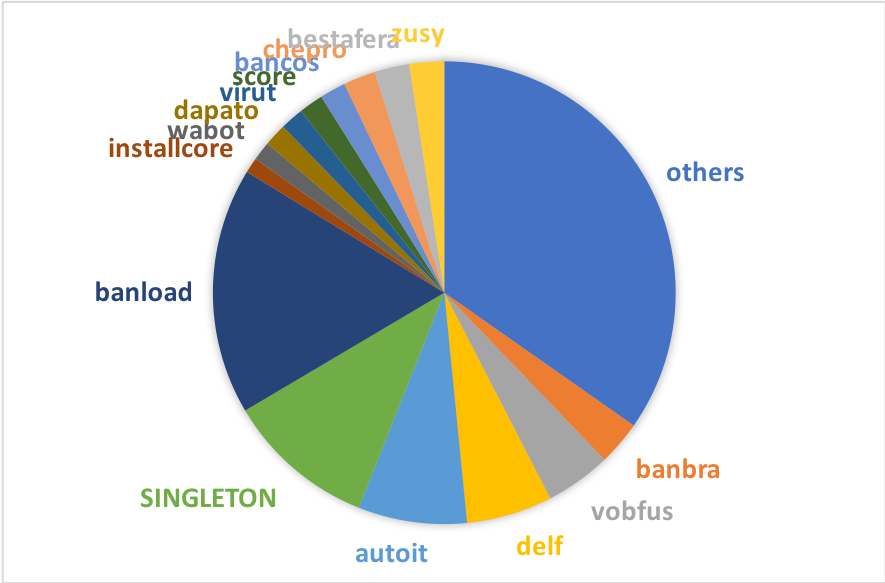}
\vspace{-0.5em}
\caption{Malware families in our dataset classified by AVClass. 
}
\label{fig:malware_dist}
\vspace{-1.5em}
\end{figure}

For the malicious part, the dataset consisted of 9,115 Microsoft Windows PE32 format malware samples collected between 2014 and 2018 from threats detected by the incident response team of a major financial institution (who chose to remain anonymous) in the corporate email and Internet access links of the institution's employees, and 7 APTs collected from Rekings~\cite{rekings}. Figure~\ref{fig:malware_dist} shows the categorization of our malware samples using AVClass~\cite{sebastian2016avclass}\footnote{AVClass is an automatic, vendor-agnostic malware labeling tool. It provides a fine-grained family name rather than a generic label (e.g., Trojan, Virus, etc.), which contains scarce meaning for the malware sample.}.
The goal of malware collection is to be as broad as possible. The distribution of our collected malware family (Figure \ref{fig:malware_dist}) correlates to the distribution of malware detected by AV companies. 
For the benign part, our dataset was composed of 1,338 samples, including 866 Windows benchmark software, 50 system software, 11 commonly used GUI-based software, and 409 other benign software. GUI-based software (e.g., Chrome, Microsoft Office, Video Player) had keyboard and mouse operations simulated using WinAutomation~\cite{winautomation}. 




\vspace{-1em}
\subsection{System Call Dataset}
\label{sec:dataset}

For each experiment, we run malware and benign software and collect system-wide system calls for five minutes. The experiments were carried out under three different scenarios: (1) running one general malware, one benchmark/GUI-based/other benign software, and dozens of system software. 
(2) running two general malware, several benchmark/GUI-based/other benign software, and dozens of system software. 
(3) running one APT, one benchmark/GUI-based/other benign software, and dozens of system software.

We randomly sampled half of the malware and benign software for training and the other half for testing, to avoid overfitting~\cite{christopher2016pattern}. We collected 493,095 traces in total. During the running of one malware sample, multiple benign processes (especially system processes) would run concurrently. Therefore, the number of benign process executions we collected would be much larger than that of malicious process executions. Our classifier would end up handling imbalanced datasets, which could adversely affect the training procedure, because the classifier would be prone to learn from the majority class. 
\xy{Hence, we under-sampled the dataset by reducing the number of sliding windows from benign processes to the same as malware processes, so that the ML and DL detectors can learn from the same number of positive and negative samples.}


Window size and stride are two important hyper-parameters in sliding window methods, indicating the total and the shifting number of n-gram system calls in each process. In real-time detection, large window sizes need more time to fill up the sliding window, provide more context for DL models, and therefore achieve higher accuracy. We aim to choose the proper two hyper-parameters to ensure malware are detected accurately and in time (before completing its malicious behavior). 
In our experiments, we observed that malware can successfully infect the systems in one to two seconds, and 1,000 system calls can be invoked in one second. Therefore, we selected and compared three window size and stride pairs: (100, 50), (200, 100), (500, 250). The results showed that with pair (500, 250), \deepm needs 1 to 2 seconds to fill up the window and analyze. With pair (200, 100), \deepm needs 0.5 to 1 second. While with pair (100, 50), \deepm needs only 0.1 to 0.5 seconds. The F1-scores for \deepm are 97.02\% , 95.08\% and 94.97\% respectively. 

\xy{Although the detection performance is increased with the increase of window sizes, large window sizes indicate  longer wait time for the detector to allow the software to be executed and fill up the windows. During this long wait time, the malware are more likely to conduct malicious behaviors. 
Therefore, a proper window size needs to be carefully selected so as to balance the prediction performance and the wait time. 
Since the performance gain with respect to the increased window size is marginal (around 2\% increase in terms of F1 score from 500 to 100), while the waiting time can be reduced to 10\%-20\%, we set the window size and stride as 100 and 50, respectively, for the following evaluation.
In practice, the value of window sizes can be altered in our \spectrum framework to meet the required detection performance (e.g., a required accuracy, or an acceptable FP rate). }

\subsection{\xy{Performance Analysis of standalone ML and DL classifiers}}
\label{sec:standalone_evaluation}
We started by comparing the performance of \deepm with relevant ML and DL classifiers.
Our metrics for model performance were accuracy, precision, recall, F1 score, and false positive (FP) rate.

\begin{table*}[!tb]
\centering
\renewcommand{\arraystretch}{1.1}
\caption{\xy{Detection performance of standalone ML or DL classifiers.} \deepm achieves the best performance among all the models. Random Forest is approximately 8\% more accurate and much faster than the other machine learning models. }
\label{tab:winauto}
\vspace{-0.5em}
\begin{tabular}{@{}lllrrrrrrr@{}}
\toprule
Models & Window Size & Stride & Accuracy & Precision & Recall & F1 Score & FP Rate & \begin{tabular}{r}
Detection Time\\ with GPU (s)\end{tabular} & \begin{tabular}{r}
Detection Time\\ with CPU (s)\end{tabular} \\ \midrule
AdaBoost & \multirow{4}{*}{100} & \multirow{4}{*}{50} & 79.25\% & 78.11\% & 81.31\% & 79.67\% & 22.80\% & N/A & 0.0187 \\
\textbf{Random Forest} &  &  & \textbf{89.05\%} & \textbf{84.63\%} & \textbf{95.44\%} & \textbf{89.71\%} & \textbf{17.35\%} & \textbf{N/A} & \textbf{0.0089} \\
XGBoost &  &  & 84.78\% & 90.99\% & 77.22\% & 83.54\% & 7.66\% & N/A & 0.0116 \\
\textbf{\deepm} &  &  & \textbf{94.84\%} & \textbf{92.63\%} & \textbf{97.43\%} & \textbf{94.97\%} & \textbf{7.76\%} & \textbf{0.0383} & \textbf{1.4104} \\\hline
AdaBoost & \multirow{4}{*}{200} & \multirow{4}{*}{100} & 78.27\% & 72.84\% & 90.19\% & 80.59\% & 33.64\% & N/A & 0.0132 \\
\textbf{Random Forest} &  &  & \textbf{93.49\%} & \textbf{89.99\%} & \textbf{97.90\%} & \textbf{93.77\%} & \textbf{10.91\%} & \textbf{N/A} & \textbf{0.0063} \\
XGBoost &  &  & 70.81\% & 63.65\% & 97.08\% & 76.89\% & 70.81\% & N/A & 0.0073 \\
\textbf{\deepm} &  &  & \textbf{94.96\%} & \textbf{93.05\%} & \textbf{97.20\%} & \textbf{95.08\%} & \textbf{7.27\%} & \textbf{0.0543} & \textbf{1.3784} \\\hline
AdaBoost & \multirow{4}{*}{500} & \multirow{4}{*}{250} & 81.81\% & 79.44\% & 85.86\% & 82.52\% & 22.24\% & N/A & 0.0108 \\
\textbf{Random Forest} &  &  & \textbf{93.94\%} & \textbf{97.16\%} & \textbf{90.54\%} & \textbf{93.73\%} & \textbf{2.65\%} & \textbf{N/A} & \textbf{0.0048} \\
XGBoost &  &  & 79.19\% & 94.14\% & 62.27\% & 74.95\% & 3.88\% & N/A & 0.0062 \\
\textbf{\deepm} &  &  & \textbf{97.03\%} & \textbf{97.54\%} & \textbf{96.50\%} & \textbf{97.02\%} & \textbf{2.43\%} & \textbf{0.0797} & \textbf{1.3344} \\ \bottomrule
\end{tabular}
\vspace{-0.5em}
\end{table*}

\xy{Table~\ref{tab:winauto} shows the detection performance and execution time of using a single ML or DL model. }
\deepm achieves the best performance among all the models. With window size 500 and stride 250, \deepm achieved an accuracy of 97.03\% and a false positive rate of 2.43\%. The performance of \deepm increased with the increase in window size and stride.
Random Forest performs the second best with an accuracy of 89.05\% at window size 100 and stride 50, and an accuracy of 93.94\% at window size 500 and stride 250, approximately 10\% better than AdaBoost and 5\% better than XGBoost. Random Forest also achieves the fastest speed among all the machine learning methods. 
While \deepm model is 3.09\% more accurate than Random Forest, its classification time is approximately 100 times slower on CPU and 3 to 5 times slower on GPU. 

We did not apply GPU devices to ML models because their classification time with conventional CPUs is much smaller (at least one order of magnitude) than those measured for DL algorithms. We denote `N/A' as not applicable in Table~\ref{tab:winauto}. In real life, GPU devices, especially specific DL GPUs for accelerating DL training/testing are still not sufficiently widespread in end-user or corporate devices.
\xy{In addition, the preprocessing time of Reconstruction Module is around 1.098 seconds per run or 0.0003 seconds per window (when using 100 window size), which is almost negligible compared with ML (0.0088 seconds per window) or DL execution time (0.0383 seconds per window using GPU or 1.4104 seconds per window using CPU).}

We observed that malware could successfully infect the target system in one to two seconds, and 1,000 system calls can be invoked in one second. Moreover, traces from one process may take a significant amount of time to fill up one sliding window because the process might be invoking system calls at a low rate. Thus, in the remainder of this section, we chose to compare all the algorithms using window sizes of 100 and strides of 50 system calls. 
For all window sizes and strides, the best ML model was Random Forest. Therefore, for \spectrum, we chose Random Forest for ML classification and \deepm for DL classification. 

In addition, we evaluated \spectrum's performance on seven APTs with various borderline intervals. \spectrum successfully detected all the APTs and six of them were subjected to \deepm classification. For different borderline intervals, the false positive rate was approximately 10\%. Random Forest in isolation detected one APT with a 52.5\% false positive rate.

\begin{table*}[!tb]
\centering
\renewcommand{\arraystretch}{1.1}
\caption{\xy{Comparison on different borderline policies for the \hybridm combining ML and DL classifiers. Borderline intervals are described with a lower bound and an upper bound. The move percentage represents the percentage of software in the system that received a borderline classification with Random Forest (according to the borderline interval) and was subjected to further analysis with \deepm.}}
\label{tab:hybrid}
\vspace{-0.5em}
\begin{tabular}{@{}llrrrrrrrr@{}}
\toprule
\begin{tabular}{l}Lower\\Bound\end{tabular} & \begin{tabular}{l}Upper\\Bound\end{tabular} & Accuracy & Precision & Recall & F1 Score & FP Rate & \begin{tabular}{r}
Detection Time\\ with GPU (s)\end{tabular} & \begin{tabular}{r}
Detection Time\\ with CPU (s)\end{tabular} & \begin{tabular}{r}Move\\Percentage\end{tabular} \\ \midrule
10\% & 90\% & 94.71\% & 92.41\% & 97.43\% & 94.85\% & 8.00\% & 0.0223 & 0.381 & 55.95\% \\
20\% & 80\% & 94.61\% & 92.23\% & 97.43\% & 94.76\% & 8.21\% & 0.0173 & 0.1543 & 48.32\% \\
30\% & 70\% & 94.34\% & 91.77\% & 97.43\% & 94.51\% & 8.75\% & 0.0146 & 0.0884 & 41.45\% \\
40\% & 60\% & 93.72\% & 90.79\% & 97.37\% & 93.96\% & 9.92\% & 0.0123 & 0.0497 & 34.96\% \\ \bottomrule
\end{tabular}
\vspace{-1.5em}
\end{table*}

\subsection{Performance Analysis of \spectrum, combining ML and DL classifiers} \label{sec:eval_spectrum}
Next, we evaluated \spectrum's performance for different borderline intervals. 
\xy{Based on the results from Table~\ref{tab:winauto}, we chose Random Forest as ML classifier and \deepm as DL classifier with window size 100 and stride 50. \spectrum is evaluated based on various borderline intervals. }
As discussed before, in \spectrum, if a piece of software receives a malware classification probability from Random Forest smaller than the lower bound, it is considered benign software. If the probability is greater than the upper bound,  it is considered malicious. However, if the classification falls within the borderline interval, the software is subjected to \deepm. \xy{Table~\ref{tab:hybrid} shows the performance of \spectrum using various (configurable) borderline intervals when combining ML and DL methods.} We chose lower bound as 10\%, 20\%, 30\%, and 40\%, upper bound as 60\%, 70\%, 80\%, and 90\%. 
\xy{We observe that when a small borderline interval is configured (e.g., 30\%-70\%, 40\%-60\%), the detection time is reduced to less than 0.1 seconds even using CPU, with minimal degradation of prediction performance.}
With a borderline interval of [30\%-70\%], \spectrum achieved an accuracy of 94.34\% and a false positive rate of 8.75\%, with 41.45\% of the samples moved for \deepm analysis.
This highlights the potential of \spectrum: approximately 60\% of the samples were quickly classified with high accuracy as malware or benign software using an initial triage with faster Random Forest. 
Only 40\% of the samples needed to be subjected to a more expensive analysis using \deepm.
\spectrum achieves high prediction performance while making it feasible to use expensive DL algorithms in real-time malware detection.

\vspace{-0.5em}
\subsection{Comparison with Other DL Malware Detectors}
\label{sec:dl_compare}
We compared \spectrum with other DL-based malware detectors in the literature leveraging system/API calls as features. 
\xyn{The features used in our work and~\cite{kolosnjaji2016deep} are Windows system calls. The features used in~\cite{hardy2016dl4md} and~\cite{wang2017adversary} are Windows API calls. System calls provide more insights on the software at the kernel level, compared with API calls collected at the user level. Moreover, we evaluated the detection performance on a much larger dataset with more traces, which provides more accurate evaluation results.}

In Table~\ref{tab:dlcomp}, we summarize the obtained results.  
\xy{Compared with the existing state-of-the-art DL-based malware detectors, our approaches achieve much better detection performance in terms of accuracy, precision, and recall.}
\xy{More importantly, current literature mainly focuses on offline analysis \cite{hardy2016dl4md,wang2017adversary,kolosnjaji2016deep} and lacks analysis of the detection time in real-time. We evaluated \deepm on a much larger dataset with better prediction performance and detection time guarantee. }

\begin{table*}[!tb]
\centering
\renewcommand{\arraystretch}{1.1}
\caption{Comparison among \deepm and other DL-based malware detection in the literature. 
} 
\label{tab:dlcomp}
\vspace{-0.5em}
\begin{tabular}{cccc}
\toprule
Work   & Model   & Dataset  & Performance  \\ \midrule
DMIN'16 \cite{hardy2016dl4md} & DL4MD   & 45,000 traces  &  Accuracy 95.65\%, Precision 95.46\%, Recall 95.8\%  \\ 
KDD'17 \cite{wang2017adversary}   & DNN  & 26,078 traces  &  Accuracy 93.99\% \footnotemark\\ 
LNCS'16 \cite{kolosnjaji2016deep} & LSTM & 4,753 traces   & Accuracy 89.4\%,  Precision 85.6\%, Recall 89.4\%  \\ 
\spectrum  & \deepm & 493,095 traces & Accuracy 97.0\%, Precision 97.5\%, Recall 97.0\% \\ \bottomrule
\end{tabular}
\vspace{-1.5em}
\end{table*}

\section{Discussion}
\label{sec:sumdis}


We proposed a real-time malware detection framework, \spectrum, accomplishing the fast speed of ML and the high accuracy of emerging DL models. Only software receiving borderline classification from an ML detector needed further analysis by \deepm, which saved computational resources, shortened the detection time, and improved accuracy.
The key contribution of \spectrum is the intelligent combination of ML with emerging modern DL methods in an effective real-time detector, which can meet the needs of high accuracy, low false-positive rate, and short detection time required to counter the next generation of malware. In this section, we discuss several challenges in the current design and potential directions of improving \spectrum in future work.
\vspace{-1em}
\noindent
\xy{\subsection{Recurrent Loop in \spectrum}} \spectrum can run into a worst-case scenario, which is having a process continuously loop between Random Forest and \deepm. For example, consider a process receiving a borderline classification from Random Forest, and then being moved to \deepm. Then \deepm classifies it as benign, and the software would continue being analyzed by Random Forest, which again provides a borderline classification for the process, and so on. We plan to mitigate such recurrent processes by combining the previous prediction of \deepm with the prediction of Random Forest in the first stage.

\noindent
\xy{\subsection{Adversarial Attacks against Malware Detection} 
\label{sec:adversarial}
Despite initial successes on malware classification, a resourceful and motivated adversary can bypass the protection mechanisms using evasion attacks. A plethora of recent studies have demonstrated that ML and DL models are vulnerable to evasion attacks, such as \textit{adversarial examples}~\cite{carlini2017towards,grosse2016adversarial,yuan2019adversarial,zhang2019adversarial}.
By adding a small perturbation to the input features of machine learning models, adversaries can mislead the detection system to classify a malware as a benign software~\cite{kreuk2018deceiving,hu2018black,liu2019atmpa,park2019generation,chen2019adversarial,suciu2019exploring,ebrahimi2020binary,yuan2020black}.
However, most adversarial attacks target at static malware detection: malware is not executed and machine learning conducts detection on the static code. For example, Park etal. injected dummy code into malware source code using an obfuscation technique~\cite{park2019generation}. Liu etal. manipulated the  malware images generated from malware binaries using adversarial attacks~\cite{liu2019atmpa}. Only~\cite{kreuk2018adversarial} generated adversarial sequences to attack dynamic behavior-based malware detection systems. However, their victim detection model is much simpler than our proposed model.
In our work, we propose a framework for dynamic malware detection, where malware is investigated during execution, which makes the adversarial attacks much easier to detect, since it is challenging to generate a sequence of system/API calls with normal behaviors. 
In the future, we plan to thoroughly investigate the robustness of \spectrum against adversarial attacks. 
}
\vspace{-1em}

\footnotetext{\xyn{The performance in terms of precision and recall is not provided in~\cite{wang2017adversary}.}}

\noindent
\xy{\subsection{Weakly Supervised Anomaly Detection} \label{sec:deep_anomaly}
In our work, we assume that the malware samples are well annotated and ML and DL models are trained on these samples, i.e., fully supervised anomaly detection. However, the malware samples have been generated and evolved rapidly in recent years. It becomes impractical to analyze and identify all the malware samples for model training. Therefore, a weakly-supervised learning paradigm~\cite{zhou2018brief} is urgently needed for malware detection, where only a limited number of labeled positive samples are provided and a large number of positive samples are not labeled.  Although unsupervised learning approaches leveraging Autoencoders and generative adversarial networks (GANs) can be used in weakly supervised learning by learning features that are robust to small deviations in negative samples~\cite{vincent2010stacked,ding2019deep,schlegl2017unsupervised,akcay2018ganomaly,zenati2018adversarially,zhou2021feature}, the labeled positive samples are not fully utilized. Recently, Pang et al. proposed DevNet and PRO to learn anomaly scores using deep learning models for better use of labeled positive data~\cite{pang2019deep,pang2019deepa}. The proposed framework in \spectrum can be further extended with the capabilities of learning from weakly labeled samples using these advanced weakly-supervised approaches.}

\section{Related Work}\label{sec:related}
Our work pertains to fields related to behavior-based malware detection. In this section, we summarize the state-of-the-art in these areas and highlights topics currently under-studied.

\vspace{-0.5em}
\subsection{Behavior-based Malware Detection}
Dynamic behavior-based malware detection~\cite{semantics-aware,lanzi2010accessminer} evolved from Forrest et al.'s seminal work~\cite{forrest96} on detecting anomalies using system
calls as features. 
Christodorescu et al.~\cite{semantics-aware,miningspec} extract high-level and unique behavior from the disassembled binary to detect malware and its variants. The detection is based on predefined templates covering potentially malicious behaviors, such as mass-mailing and unpacking. 
Willems et al.~proposed CWSandbox, a dynamic analysis system that monitors malware API calls in a virtual environment~\cite{willems2007cwsandbox}. Rieck et al.~\cite{rieck2008learning} leveraged API calls as features to classify malware into families using Support Vector Machines (SVM), and processed system calls into q-grams representations using an algorithm similar to the k-nearest neighbors (kNN)~\cite{rieck2011automatic}. Mohaisen et al.~\cite{mohaisen2015amal} introduced AMAL, a framework to dynamically analyze and classify malware using SVM, linear regression, classification trees, and kNN. Kolosnjaji et al.~\cite{Kolosnjaji2016adaptive} proposed maximum-a-posteriori (MAP) to classify malware families using Cuckoo's Sandbox~\cite{cuckoo}. Although these techniques were successfully applied for malware classification, they did not consider benign samples~\cite{bailey2007automated}, and were limited to labeling unknown malware pertaining to one of the existing clusters. 

Wressnegger et al.~\cite{wressnegger2016comprehensive} proposed Gordon for detecting Flash-based malware. Gordon considered execution behavior from benign and malicious Flash applications and generated \textit{n-grams} for an SVM model. 
Bayer et al.~used a modified version of QEMU to monitor API calls and breakpoints~\cite{bayer2006ttanalyze}. This approach was later used to build Anubis~\cite{anubis}. Anubis is better suited for offline malware analysis (by providing a detailed execution report). \spectrum, on the other hand, focuses on real-time detection and monitors a more diverse set of system calls. 
Kirat et al.~introduced BareBox, which hooked system calls directly from SSDT~\cite{kirat2011barebox}. Barebox runs in bare metal, and can potentially obtain behavioral traces from malware equipped with anti-analysis techniques. Barebox aims in live system restoring and analyzed on only 42 malware samples. \spectrum also uses system call hooking to monitor software behavior, but is larger scale and uses a testbed to run malware automatically. 
Anubis \cite{anubis} works best for offline malware analyzers by providing a detailed execution report while \spectrum
delivers on-thy-fly detection results to end users. \spectrum improves the experiment of Anubis by monitoring more kinds of system calls, running a longer time for each experiment, and using cutting-edge DL models to help with the detection.

Some lines of work focus on developing tools to detect evasive malware~\cite{balzarotti2010efficient}. \spectrum is  resilient to anti-analysis because its monitoring mechanism operates at the kernel level. Other works using tainting techniques, e.g., VMScope~\cite{johnson2011differential}, TTAnalyze \cite{bayer2006ttanalyze}, and Panorama \cite{yin2007panorama}, emulate environments with QEMU, and trace the data-flow from whole-system processes. \spectrum differs from such works by monitoring software behavior through system-wide system call hooking instead of data tainting, while maintaining the interactions among different processes in a lightweight manner. Contrary to \spectrum, such tools might encounter challenges when deployed in practice. For example, Panorama \cite{yin2007panorama} relies on a human to manually label address and control dependency tags that should be propagated.
Canali et al.~evaluated different types of models for malware detection. Their findings confirm that the accuracy of some widely used learning models is poor, independently of the values of their parameters, and high-level atoms with arguments model performs the best, which corroborates our experimental results. Further, the paper points out that on-the-fly detection suffers from even higher false-positive rates because of the diversity of applications and the system calls invoked~\cite{lanzi2010accessminer}. All these findings corroborate the results of our paper.

\vspace{-1em}
\subsection{ML-based Malware Detection} 
Xie et al.~proposed a one-class SVM model with different kernel functions~\cite{xie2014evaluating} to classify anomalous system calls in the ADFA-LD dataset~\cite{creech2013generation}.
Ye et al.~proposed a semi-parametric classification model for combining file content and file relation information to improve the performance of file sample classification~\cite{ye2011combining}. Abed et al.~used bags of system calls to detect malicious applications in Linux containers~\cite{abed2015applying}. Kumar et al. used K-means clustering~\cite{kumar2013k} to differentiate legitimate and malicious behaviors based on the NSL-KDD dataset. Fan et al.~used a sequence mining algorithm to discover malicious sequential patterns and trained an All-Nearest-Neighbor (ANN) classifier based on these discovered patterns for malware detection~\cite{fan2016malicious}.

The Random Forest algorithm has been applied to classification problems as diverse as offensive tweets, malware detection, de-anonymization, suspicious Web pages, and unsolicited email messages~\cite{Chatzakou2017,Mariconti2017}.
Conventional ML-based malware detectors suffer, however, from high false-positive rates because of the diverse nature of system calls invoked by applications, as well as the diversity of applications~\cite{lanzi2010accessminer}.
\vspace{-1em}

\subsection{DL-based Malware Detection} 
There are recent efforts to apply DL for malware detection. 
Deep learning has made great successes in speech recognition~\cite{amodei2016deep}, language translation~\cite{sutskever2014sequence}, speech synthesis~\cite{dieleman2016wavenet}, and other sequential data~\cite{schmidhuber2015deep}.
Li et al. ~proposed a hybrid malicious code classification model based on AutoEncoder and DBN and acquired a  relatively high classification rate on the part of the now outdated KDD99 dataset~\cite{li2015hybrid}. 
Pascanu et al.~first applied deep neural networks (recurrent neural networks and echo state networks) to model the sequential behaviors of malware. They collected API system call sequences and C run-time library calls and detected malware as a binary classification problem~\cite{pascanu2015malware}. David et al.~\cite{david2015deepsign} used a deep belief network with denoising autoencoders to automatically generate malware signatures for classification. Saxe and Berlin~\cite{saxe2015deep} proposed a DL-based malware detection technique with two-dimensional binary program features and provided a Bayesian model to calibrate detection. 
Huang and Stokes ~\cite{huang2016mtnet} designed a neural network to extract high-level features and classify them into both benign/malicious and also 100 malware families based on shared features. 
Dahl et al.~detected malware files by neural networks and used random projections to reduce the dimension of sparse input features~\cite{dahl2013large}. 
Wang ~\cite{wang2016droiddeeplearner} extracted features from Android Manifest and API functions to be the input of the deep learning classifier. 
Hou et al.~collected the system calls from the kernel and then constructed the weighted directed graphs and use DL framework to make dimension reduction~\cite{hou2016deep4maldroid}. 
All these DL-based methods rely on handcrafted features. 
Recently, Kolosnjaji et al.~\cite{kolosnjaji2016deep} proposed a DL method to detect and predict malware families based on system call sequences. 
Our proposed \deepm has shown superior performance over their approaches by coping the spatial-temporal features with ResNext designs.
To cope with the fast-evolving nature of malware, Transcend \cite{jordaney2017transcend} addresses concept drift in malware classification. 
Transcend can detect aging machine learning models before their degradation. As future work, we plan to leverage Transcend's concept drift model in \spectrum for re-training.


Our work addressed the performance of DL-based malware detection algorithms running in a real system.
Compared with the recent work (\cite{hardy2016dl4md,wang2017adversary,kolosnjaji2016deep}), our experiments show that deep learning performs fewer false-positive samples compared with conventional machine learning models. Meanwhile, conventional machine learning algorithms run about 2 to 3 orders of magnitude faster than deep learning ones.

\vspace{-0.5em}
\section{Conclusion}
\label{sec:conclusion}


In this paper, we introduced \spectrum, a novel paradigm and framework for real-time malware detection, which combines the best of conventional machine learning and deep learning techniques using multi-stage classification.
In \spectrum, all software in the system start execution subjected to a conventional machine learning detector for fast classification. If a piece of software receives a borderline classification, it is subjected to further analysis via more performance expensive and more accurate deep learning methods, via our novel algorithm \deepm. 
\deepm utilizes both process-wide and system-wide system calls and predicts malware in both short and long-term ways.
With a borderline interval of [30\%-70\%], \spectrum achieved an accuracy of 94.34\% and a false positive rate of 8.75\%, with 41.45\% of the samples moved for \deepm analysis. The detection time using GPU is 0.0146 seconds on average. Even using CPU, the detection time is less than 0.1 seconds.



On one hand, our work provided evidence that conventional machine learning and emerging deep learning methods in isolation might be inadequate to provide both high accuracy and short detection time required for real-time malware detection. 
On the other hand, our proposed \spectrum combines the best of ML and DL methods and has the potential to change the design of the next generation of practical real-time malware detectors.

\vspace{-0.5em}
\section*{Acknowledgment}
This material is based upon work supported by the National Science Foundation under Grant No 1801599. This material is based upon work supported by (while serving at) the National Science Foundation.
\vspace{-0.5em}
\bibliographystyle{IEEEtran}
\bibliography{sample.bib,biblio.bib,ml.bib,adv.bib}


\end{document}